\documentclass{article}

\usepackage{graphicx}
\usepackage{amsmath}
\usepackage{listings}
\usepackage{comment}
\usepackage{CJKutf8}
\usepackage{amsfonts}
\usepackage{amssymb}
\usepackage[utf8]{inputenc}
\usepackage{booktabs}
\usepackage[table]{xcolor}
\usepackage{float}
\usepackage{setspace}
\usepackage{mathtools}
\usepackage{caption}
\usepackage{microtype}
\usepackage{subcaption}
\usepackage{authblk}
\usepackage[bottom]{footmisc}
\usepackage{float}
\usepackage[backend=biber, style=authoryear, sorting=nyt, uniquename = false]{biblatex}
\usepackage[margin=1in]{geometry}

\title{\texttt{GephiForR}: An R package for creating Gephi-style network visualizations}
\author{Julia Manso\thanks{Correspondence: Julia Manso, Department of Statistics, 24-29 St Giles', Oxford OX1 3LB, United Kingdom. \newline Email: jumanso@stats.ox.ac.uk}}
\affil{\small{\textit{Department of Statistics, University of Oxford, Oxford, U.K.}}}

\date{}

\begin{document}

\maketitle
\begin{abstract}
This paper introduces \texttt{GephiForR}, an R package designed to replicate Java-based Gephi's key plotting tools in R. The package is accessible to those with minimal R experience and, in particular, implements ForceAtlas2, the key layout feature developed for Gephi by Jacomy et al. (2014). \nocite{jacomy_forceatlas2_2014} The most significant advancement is the ability to pass previous positions into ForceAtlas2 as baselines, a particularly useful feature for plotting the evolution of network layouts for time series data. \texttt{GephiForR} is especially suited for networks of less than 1000 nodes, simply because R's dependence on single-thread computation means that larger networks take longer to compute, but the package can handle these larger networks as well. I demonstrate the package's capabilities through various examples and comparisons with existing tools and Gephi itself, assessing performance and speed. 
\end{abstract}
\vspace{0.5cm}

\noindent \textbf{Keywords\textemdash} \texttt{GephiForR}, network visualization, time series, Gephi, R package

\section{Introduction}

A popular visualization tool for plotting networks, Gephi is a Java-based software created by a team of students at the University of Technology of Compiègne in 2008. The software allows for easy network manipulation, for instance, by allowing a user to run a layout algorithm, rotate nodes, and expand or contract the network. Accessible to those without a strong coding background and able to handle large networks (including those with over 1 million nodes), Gephi also contains custom-built layout algorithms like the standard ForceAtlas2 (\cite{jacomy_forceatlas2_2014}) and ForceAtlas2 implementing Noack (2009)'s LinLog model, \nocite{noack_modularity_2009} both of which offer force-based approaches to network visualization. Yet, Gephi has been plagued by issues with Java versions, making it difficult for users to install and use, and most critically, its file type ``.gephi" is unstable. Many users have reported that ``.gephi" files are easily corrupted and unable to be opened after they are saved (\cite{jacomy_is_2018}). Replication then becomes challenging:  as Gephi randomly assigns initial positions to nodes, running the same series of layout transformations on the same data can produce a network that looks very different. Further, Gephi is limited in its ability to illustrate the evolution of a network over time (i.e., in time series data), as it does not dynamically update layout position across periods.\footnote{This is particularly the case for a fully connected network, as Gephi uses node degree as the basis for showing network change over time. Displaying a time series with no new nodes or edges, then, results in a static layout, regardless of whether periods are marked by either intervals or timestamps.}

In light of these shortcomings, I create the \texttt{GephiForR} package to streamline integration with broader \texttt{igraph} objects and facilitate the visualization of time series data. Designed to be accessible to those without an extensive R background, \texttt{GephiForR} allows users to generate Gephi-like graphs with relative ease. Most notably, \texttt{GephiForR} allows a layout to be passed in as the initial position for a new period, enabling more stable node positions between sequential periods and the easy replication of plots. The package implements ForceAtlas2, with options for LinLog and stronger gravity, as well as functions for scaling and rotating graphs. I also include functions that facilitate easy plotting, such as node and edge coloring by attribute, as well as a Gephi-style \texttt{igraph} plotter that requires minimal adjustments. The package can be downloaded from the Comprehensive R Archive Network (CRAN) at \url{https://CRAN.R-project.org/package=GephiForR}.

In the following section, I outline the statistical tools implemented in \texttt{GephiForR}. Then, in Section \ref{package overview}, I describe how to use the package and provide examples utilizing the functions. Next, in Section \ref{Comparison with Existing Tools}, I compare plots generated with \texttt{GephiForR} to those generated by Gephi, as well as by another attempt to bring the ForceAtlas2 algorithm to R. After evaluating efficiency and scale in Section \ref{efficiency and scale}, I conclude with a discussion in Section \ref{conclusion}.

\section{Methods} \label{methods}
\subsection{ForceAtlas2: Base Case}  \label{FA2 base}
This section draws heavily on Jacomy et al. (2014) as I follow their implementation and explanation of ForceAtlas2. Specifically, ForceAtlas2 calculates the ``net forces" acting on each node by summing the node's attraction to and repulsion from each other node it connects to (i.e., shares an edge with). The formulas for both attraction and repulsion are rooted in Eades (1984) \nocite{eades_heuristic_1984} and his application of physics principles to networks; the attraction formula is an altered version of Hooke's law, which reflects the compression behavior of a spring ($F = -k \times x)$, where $F$ is the spring force, $k$ is the spring constant in units of Newtons/meter, and $x$ is the spring compression. Repulsion is based on Coulomb's law, which calculates forces between electrically charged particles ($F = k\frac{n_1n_2}{d(n_1, n_2)^2}$). Here, $F$ is the resulting force; $n_1$ and $n_2$ are the point charges of particles 1 and 2, respectively; $d(n_1, n_2)$ is the distance between the two particles $n_1$ and $n_2$; and $k$ is the Coulomb's law constant ($9.0 \times 10^9\textit{ }Nm^2/C^2$).\footnote{The units here are Newtons times meters squared over Coulombs squared.}
This combination of ``spring-like" attractive forces and ``particle-like" repulsive forces has persisted, with several authors offering slightly modified equations over the past 40 years.

The combination of equations used in Gephi was derived by Gephi's creators (\cite{jacomy_forceatlas2_2014}), and they build upon this ``spring-particle" model, as well as Noack (2007)'s ideas on force-directed algorithms for node placement. \nocite{noack_energy_2007} Noack posited that the most significant difference among force-directed algorithms is the role played by distance, with the interdependence between distance and forces being linear, exponential, or logarithmic (\cite{jacomy_forceatlas2_2014}; \cite{noack_energy_2007}). He defines the energy model or ``(attraction, repulsion) model" of a layout as the exponent taken by distance in the formulas used to calculate both attraction and repulsion, respectively, with log being considered the $0^{\text{th}}$ power (\cite{noack_modularity_2009}). This means that the attractive and repulsive force combination of the ``spring-particle" model is (1, -2). Gephi is unique because it uses (1, -1), unlike Noack's algorithm LinLog (0, -1), the ``spring-particle" model (1, -2), or another very common layout algorithm, Fruchterman and Reingold (2, -1) (\cite{noack_modularity_2009}).\footnote{In their paper, Fruchterman and Reingold (1991) define attraction and repulsion to be $F_a = d^2/S$ and $F_r = -S^2/d$, respectively, where $d$ is distance and $S$ is a scaling factor. It thus has an (attraction, repulsion) form of (2, -1). \nocite{fruchterman_graph_1991}}

In this framework, ``distances are less dependent on densities for large [attraction] - [repulsion values], and less dependent on path lengths for small [attraction values]" (\cite{noack_modularity_2009}, p. 3). This means, for instance, that ForceAtlas2 is better at showing clusters compared to more traditional node position algorithms like that of Fruchterman and Reingold (1991) \nocite{fruchterman_graph_1991} but is less efficient than a method like LinLog.\footnote{Note that in line with Gephi, I build a LinLog option in ForceAtlas2 which uses the alternate attractive formula described in Section \ref{ForceAtlas2 LinLog Mode}.} 

In Gephi, repulsion $F_r$ is calculated by the following formula, which is a slightly modified version of Coulomb's law where node degree takes the place of point charges, the constant becomes scalable rather than fixed, and distance is no longer squared. The key development from previous node repulsion calculations was the addition of ``+1" to the degree, as Jacomy et al. wanted to ensure that nodes with degrees of zero still have repulsive force. The formula is thus 
\begin{align}
    F_{r} = \frac{S(deg(n_1)+1) (deg(n_2)+1)}{d(n_1, n_2)}
    \label{F_r eq}.
\end{align}
This calculation is repeated between every possible node pair in the data. Thus, $d(n_1, n_2)$ is the distance between the two nodes $n_1$ and $n_2$ in the pair, $deg(n_1)$ and $deg(n_2)$ are the degrees of each of the two nodes, and $S$ is a scalable constant that influences the repulsion level in the graph, with higher repulsion making a sparser graph. Gephi sets the baseline built-in value of $S$ as 10. 

The attractive force $F_{a}$ is the product of the edge weight $w(e)^\delta$ between each pair of nodes, multiplied by the geometric distance between them, as shown below:
\begin{align}
    F_{a} = w(e)^\delta d(n_1, n_2),
    \label{attraction formula}
\end{align}
where $\delta$ is the binary variable ``Edge Weight Influence," set to either 0 or 1. Unlike Hooke's law, this formula does not have a constant; Jacomy et al. reason that if a constant were included in equation (\ref{attraction formula}), it would work against the constant $S$ of equation (\ref{F_r eq}), playing the opposite role in the spatialization of the graph. They thus dropped the constant from the attractive equation so that $S$ of equation (\ref{F_r eq}) is the only constant involved in scaling. Jacomy et al. instead included edge weights $w(e)^\delta$, as these can have a dramatic effect on the attractive force between two nodes. Including edge weights allows the attractive force to reflect the strength of relationships between nodes, but Gephi allows a user to ignore edge weights if desired (in which case $\delta$ is set to 0). 

Both the attraction and repulsion forces can be decomposed into vector form and summed to calculate the net forces on each node; this value is then used to find the resulting displacement $\Delta(n)$ via the formula: 
\begin{align}
    \Delta(n) = s(n)*F(n)
    \label{displacement eq},
\end{align}
where $s(n)$ is the speed of node $n$, and $F(n)$ is a vector of the sum of all the forces applied to each node $n$ (such that $F = F_{r} + F_{a}$).\footnote{If a stronger gravity $F_g$ were included, it would also be included in this sum, such that $F = F_{r} + F_{a} + F_{g}$. Stronger gravity increases the attractive force in the network and draws the nodes closer together; see Section \ref{ForceAtlas2 with Stronger Gravity} for further details.}

To calculate the speed, it is first critical to understand the two key metrics that shape it: irregular movement (``swinging") and useful movement (``effective traction"). Swinging is defined as the ``divergence between the force applied to [a node] $n$ at a given step and the force applied to $n$ at the previous step" (\cite{jacomy_forceatlas2_2014}, p. 7). $swg_{(t)}(n)$\textemdash the swinging of node $n$ at time $t$\textemdash is thus calculated as the difference between the net force at time $t$ applied to node $n$ ($F_{(t)}(n)$) and the net force applied to node $n$ at time $t-1$ $(F_{(t-1)}(n))$:
\begin{align}
    swg_{(t)}(n) = |F_{(t)}(n) - F_{(t-1)}(n)|.
    \label{swinging eq}
\end{align}
As Jacomy et al. highlight, for a node moving toward its balancing position, $swg(n)$ is close to 0, but a node that experiences forces significantly different from the previous period has a high swinging value. 

Subsequently, the global swinging value $swg(G)$ can be calculated by summing the local swinging values, weighted by the degree of each node; like in equation (\ref{F_r eq}), Jacomy et al. add 1 to each node's degree to account for nodes of 0 degree. Global swinging is thus $swg(G) = \sum_n (deg(n) + 1)swg(n)$. 

On the other hand, the effective traction $tra(n)$ of a node is the amount of ``useful force" applied to that node\textemdash that is, forces that contribute to the node's convergence. It is calculated as the average of the net force applied to node $n$ at time $t$ and that applied in the previous period ($t-1$): it is therefore 
\begin{align}
    tra_{(t)}(n) = \frac{|F_{(t)}(n) + F_{(t-1)}(n)|}{2}.
    \label{traction eq}
\end{align}
This formula means that nodes who keep their course have $tra(n) = F(n)$ and those that revert to their former positions (a perfect swing) have $tra(n) = 0$. Global effective traction $tra(G)$ is the sum of local effective traction values, weighted by the degree +1 of each node, with the +1 again accounting for possible nodes of degree 0. It is therefore calculated as $tra(G) =  \sum_n (deg(n) + 1)tra(n)$. Global speed $s(G)$ ``keeps the global swinging $swg(G)$ under a certain ratio $\tau$ of the global effective traction $tra(G)$" (\cite{jacomy_forceatlas2_2014}, p. 9). It is thus defined as 
\begin{align}
   s(G) = \tau\frac{tra(G)}{swg(G)}. 
    \label{global speed eq}
\end{align} 
$\tau$, the tolerance to swinging, can be set by the user, but Gephi also includes an algorithm to calculate it based on the density of the network and its value in previous iterations.\footnote{I replicate this algorithm in my implementation of ForceAtlas2 in \texttt{GephiForR} as well.} The speed of each node is then calculated as: 
\begin{align}
    s(n) = \frac{k_ss(G)}{(1+s(G)\sqrt{swg(n)})},
\end{align}
where $k_s$ is a constant set to 1, unless the user changes it. As Jacomy et al. (2014) describe, the logic for having a local speed for each node $s(n)$ rather than just having a global speed $s(G)$ is rooted in providing more precision for nodes that struggle to converge. The global speed is as high as possible while being limited by the tolerance $\tau$; the local speed, on the other hand, can slow nodes down but cannot speed them up. Putting this behavior together means that the ``local speed regulates the swinging while the global speed [indirectly] regulates the convergence" (\cite{jacomy_forceatlas2_2014}, p. 9). 

Ultimately, the more a node swings, the slower it moves, while nodes with very little swinging move at rates near the global speed. Finally, with the speed $s(n)$ calculated, the displacement $\Delta(n)$ can be computed per equation (\ref{displacement eq}). Final node position $p$ in each iteration $r$ is thus $p_r(n) = p_{r-1}(n) + \Delta_r(n)$, and it is ultimately a reflection of the forces acting on the node and the speed at which each node moves, with the network gradually converging toward a stable position across iterations for each time period. In \texttt{GephiForR}'s implementation of ForceAtlas2, the built-in value is \verb|iterations = 100|, although users should adjust this parameter based on the number of nodes: larger networks may require more timesteps to reach convergence while smaller ones may converge more quickly.

In Gephi itself, displaying a network for each period in a time series is time-consuming and only partially informative, given that the node position from the previous period cannot be passed in. This means, for instance, that the same network could look radically different in Gephi because nodes' initial positions were randomly assigned. Thus, even after the network is formatted with ForceAtlas2, the absolute positions of the node will be different even while the distance between neighboring nodes is the same. For example, the final distance between node A and node B would be roughly the same across initial positions, but the nodes may appear in different places in the network (for instance, on the right or left side). 

Thus, one of the key modifications in my implementation of ForceAtlas2 is that it allows layout continuity across periods: while Gephi randomly assigns each node an initial position, \texttt{GephiForR}'s ForceAtlas2 function allows the beginning position for each node to be determined by a layout that has already been calculated. In practice, this means that the previous day's layout can be passed in as the initial position, so it is possible to directly observe the movement of nodes across the network continuously from day-to-day in time series data. This behavior does not affect any of the calculations described above but does have the added benefit of stabilization across periods for nodes that do not experience significant changes in $F_r$ and $F_a$\textemdash and allows for easy detection of those that do have large changes in position. It also facilitates easy re-generation of plots, as the layout can be exported and saved. 

\subsection{ForceAtlas2 with Strong Gravity} \label{ForceAtlas2 with Stronger Gravity}

The ``strong gravity" option includes an additional force (gravity) in the net forces acting upon a node, such that $F = F_{r} + F_{a} + F_{g}$. Stronger gravity increases the attractive force in the network and draws the nodes closer together, although Jacomy et al. warn that it is sometimes so strong that it can create biased node positions and highly contracted layouts. They calculate gravity for node $n$ as 
\begin{align}
    F_{g}(n) = k_g log(deg(n)+1),
    \label{gravity formula}
\end{align}
where $deg(n)$ is the degree of node $n$. Increasing the gravity parameter $k_g$, then, prevents clusters from drifting away from each other, ``[attracting] nodes to the center of the spatialization space'' (\cite{jacomy_forceatlas2_2014}, p. 4). 

Thus, enabling strong gravity alters the calculation of net forces acting on a node ($F(n)$), but the rest of the calculations discussed in Section \ref{FA2 base} remain unchanged.

\subsection{ForceAtlas2 LinLog Mode} \label{ForceAtlas2 LinLog Mode}

As described in Section \ref{FA2 base}, Noack (2007)'s LinLog model is also adapted for Gephi. LinLog is appealing because it corresponds to groupings by Newman's modularity (\cite{newman_analysis_2004}; \cite{newman_modularity_2006}). Newman observed that nodes have stronger connections with other nodes in their communities, and communities are groups with denser relationships. Newman's modularity is a measure of this proximity, defined as the ``fraction of edges that fall within communities minus the expected value of the same quantity if edges are assigned at random, conditional on the given community memberships and the degrees of vertices" (\cite{newman_analysis_2004}, p. 6). Noack's LinLog, then, is particularly appealing because it calculates layouts such that communities appear as groups of nodes. That is, the LinLog layout tends to reflect the clusters based on the underlying modularity values.

As described in Section \ref{FA2 base}, Noack's LinLog has an (attraction, repulsion) form of (0, -1), compared to Gephi's (1, -1).\footnote{Note that ForceAtlas2 tends to also display clusters in line with LinLog, which makes sense given the relative similarities in calculation methods.} In Jacomy et al.'s implementation, the repulsive formula is the same as that in equation (\ref{F_r eq}), but attraction $F_a$ is now calculated as
\begin{align}
    F_{a} = log(1+d(n_1, n_2)).
    \label{attraction formula linlog}
\end{align}

Jacomy et al. (2014) acknowledge that this formula is slightly different from Noack's original implementation since they again add 1 to the distance to ``manage superposed nodes, as log(0) would produce an error" (p. 3). Yet, this lines up with Noack's assumptions in his original definition of the LinLog model, as he assumes ``different nodes have different positions" to avoid infinite energies (2007, p. 461). 

Compared to the normal version of ForceAtlas2, LinLog mode separates clusters from other nodes, and the nodes of each cluster are close together (\cite{noack_energy_2007}). As highlighted by Jacomy et al., switching to LinLog mode requires adjustment of the scaling parameter $S$, as the introduction of the logarithmic attractive force calculation can generate clusters distant from one another. Thus, lowering the scaling parameter $S$ will make the network tighter. 

\section{Package Overview} \label{package overview}

\subsection{Installation}
The \texttt{GephiForR} package can be installed from CRAN and loaded into R using the following commands:

\begin{verbatim}
install.packages("GephiForR")
library("GephiForR")
\end{verbatim}

\subsection{Main Features and Examples} \label{main features and examples}

The package provides six main features: calculating node layout with ForceAtlas2 (\texttt{layout.forceatlas2}), scaling node positions (\texttt{scale\_node\_positions}), rotating the layout (\texttt{rotate\_layout}), assigning node colors 
\newline (\texttt{assign\_node\_colors}), assigning edge colors (\texttt{assign\_edge\_colors}), and plotting (\texttt{easyplot}). \nocite{manso_gephiforr_2024} These are the most commonly used functions of Gephi that do not yet exist in R; other popular Gephi layout tools like Fruchterman-Reingold are already implemented in R (for instance, in \texttt{igraph}'s built-in \texttt{layout\_with\_fr} function).  

\subsubsection{\texttt{layout.forceatlas2}} \label{FA2 main feature}
\texttt{layout.forceatlas2} takes as input an \texttt{igraph} object and calculates the network's layout with the ForceAtlas2 algorithm outlined in Section \ref{FA2 base}. There are several parameters that can be adjusted depending on the features of the data. \texttt{iterations} is set to 100 by default, but smaller networks may require fewer iterations while larger ones will require more. If \texttt{plotstep} is set to a non-zero number, \texttt{layout.forceatlas2} will print intermediate plots in the console, displaying node movement between the iterations at intervals of the step size. The default is \texttt{plotstep = 10}, and labels for each node in the plot are included (\texttt{plotlabels = TRUE}). 

The function also includes options for the other layout algorithms described in Sections \ref{ForceAtlas2 with Stronger Gravity} and
\ref{ForceAtlas2 LinLog Mode}, with binaries for \texttt{linlog} and \texttt{stronggravity}. Setting each to \texttt{TRUE} enables each mode, and adjusting \texttt{gravity} changes the strength of the gravity force.\footnote{The \texttt{gravity} factor is only activated when \texttt{stronggravity = TRUE}.} 

Furthermore, as described, the function also has a \texttt{pos} argument, wherein a matrix of initial positions can be passed in, and it can also have a set center of gravity, if desired, with the parameter \texttt{center}. The only other adjustable inputs are \texttt{jittertol}, which influences the auto-calculated $\tau$ value of equation (\ref{global speed eq}) to capture how much swinging is allowed. This parameter is set to 1 by default, and Jacomy et al. do not recommend increasing it above 1. 

Throughout this paper, I include an example using the bank volatility data of Demirer et al. (2018) \nocite{demirer_estimating_2018} who estimate global banking connectedness before and after the 2008 financial crisis. After estimating a vector autoregressive (VAR) model with an adaptive elastic net on volatility data for 96 global banks, Demirer et al. use generalized forecast error variance decompositions to quantify the spillover effects among these banks. These generalized forecast error variance decompositions are subsequently used as edge weights in the network, with each node reflecting each bank. I thus replicate their network plots, generated in Gephi, with \texttt{GephiForR} for comparison. 

The step-by-step implementation, beginning with the calculation of ForceAtlas2, is as follows.

\begin{figure}[H]
    \centering
    \includegraphics[width=1.01\linewidth]{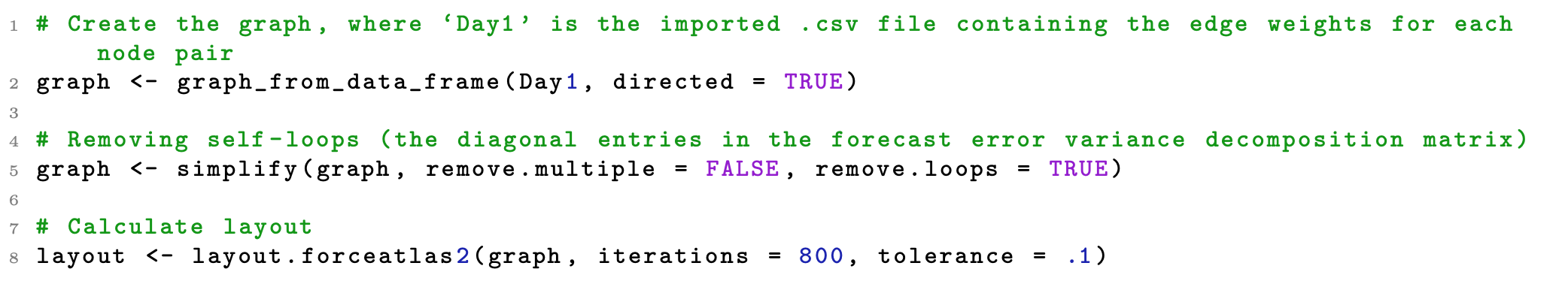}
\end{figure}

As described, this generates a two-column matrix with $N$ rows, where $N$ is the number of nodes. Importantly, because ForceAtlas2 randomly assigns initial positions, running this code two separate times will not generate the same layout result unless a seed is set (e.g., \texttt{set.seed(10)}). 

In a time series setting, an initial layout can be passed into \texttt{layout.forceatlas2} (i.e., as \texttt{pos = previous.layout}) to allow for continuity in node position across sequential periods. An example is included below. 

\begin{figure}[H]
    \centering
    \includegraphics[width=1.08\linewidth]{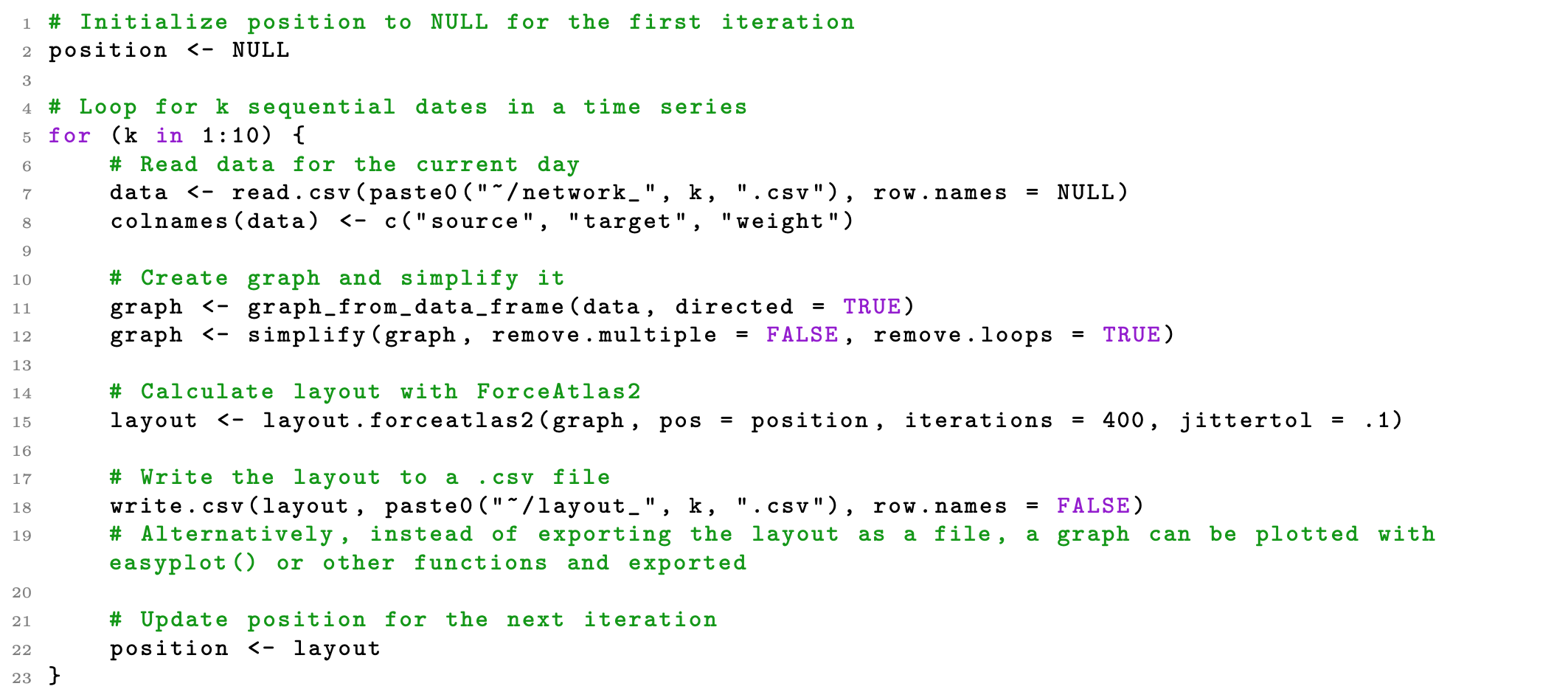}
\end{figure}

\subsubsection{\texttt{scale\_node\_positions}}

After calculating the layout, the \texttt{scale\_node\_positions} function can be used to scale node positions to facilitate visualization. It takes as input the $N \times 2$ layout matrix and the scale factor, and it returns an adjusted $N \times 2$ layout position matrix. 

\begin{figure}[H]
    \centering
    \includegraphics[width=1.05\linewidth]{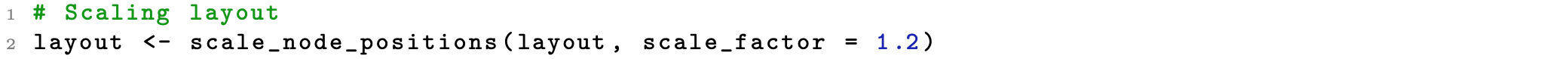}
\end{figure}

Note that it can sometimes be difficult to visualize the impact of the scaling because R's network plots are often unscaled (including that in \texttt{easyplot}). The difference is readily apparent in scaled plots, for instance, as implemented below: 

\begin{figure}[H]
    \centering
    \includegraphics[width=1.05\linewidth]{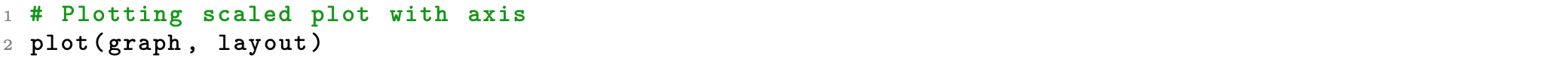}
\end{figure}

\subsubsection{\texttt{rotate\_layout}}

Next, the entire layout can be rotated for easy visualization or comparison. The \texttt{rotate\_layout} function simply takes as input the layout and angle of rotation (in degrees). As above, it produces an updated $N \times 2$ layout position matrix. 
\begin{figure}[H]
    \centering
    \includegraphics[width=1.04\linewidth]{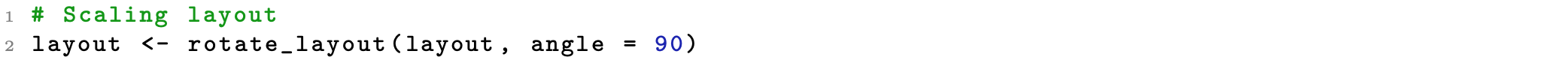}
\end{figure}

\subsubsection{\texttt{assign\_node\_colors}}

After finalizing the layout, node colors can be assigned by grouping according to the relevant feature. For instance, in the case of Demirer et al. (2018)'s data, I define an attribute for each bank's continent and assign node colors based on these continent groups. In practice, any feature of interest can be passed into the function, and coloring nodes by features is useful to illustrate clustering or underlying patterns.

\begin{figure}[H]
    \centering
    \includegraphics[width=1.05\linewidth]{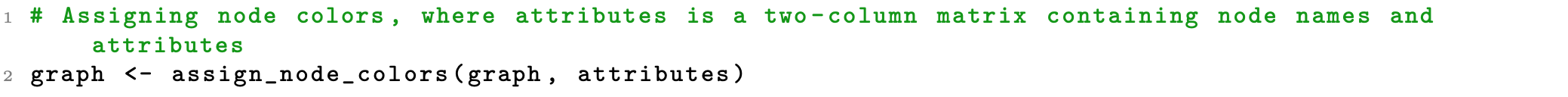}
\end{figure}

\subsubsection{\texttt{assign\_edge\_colors}}

While the \texttt{plot()} function from \texttt{graphics} formats every edge to be the same color, edge colors can also be dynamically assigned based on an edge's origin node. Doing so allows easy visualization of which nodes are connected and which direction has the stronger edge weight in a directed network. 
\begin{figure}[H]
    \centering
    \includegraphics[width=1.05\linewidth]{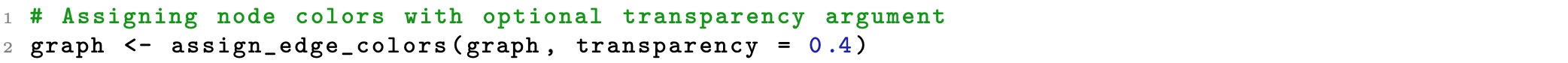}
\end{figure}

\subsubsection{\texttt{easyplot}}
Finally, bringing all of these functions together, it is simple to generate a clear R plot that mirrors Gephi. This is the goal of \texttt{easyplot}, which utilizes the \texttt{graphics::plot()} function but has many parameters specified to suppress unhelpful features of R network graphs that make visualization difficult. After passing in the \texttt{igraph} object and the $N \times 2$ layout array, the function also has arguments for label size (\texttt{label\_size}), edge color (\texttt{edge\_color}), vertex size (\texttt{vertex\_size}), edge arrow size (\texttt{edge\_arrow\_size}), and vertex label color (\texttt{vertex\_label\_color}). Several of these, including label size, edge arrow size, and vertex size, will need to be adjusted based on the size of the plots generated.\footnote{The defaults for these parameters are \texttt{label\_size = 3}, \texttt{vertex\_size = rep(3, vcount(graph))}, and \texttt{edge\_arrow\_size = 0.2}, \texttt{vertex\_label\_color = "black"}, and \texttt{edge\_color = NULL}.} 

The following is an example, with an export function wrapper to export the image as a clear JPEG file. 

\begin{figure}[H]
    \centering
    \includegraphics[width=1.05\linewidth]{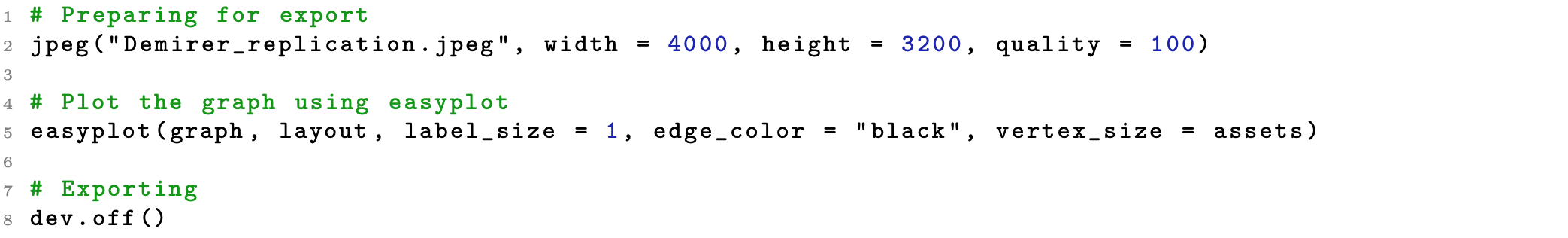}
\end{figure}

Note that the \texttt{vertex\_size = assets} argument is setting node size to be the bank's amount of assets, in line with Demirer et al. (2018). The \texttt{easyplot} function is able to handle any numeric vector of the same length as the number of vertices, and it is thus easy and convenient to adapt node size based on relevant attributes.

\section{Comparison with Existing Tools} \label{Comparison with Existing Tools}
\begin{figure}[H]
  \caption{Comparing Gephi's ForceAtlas2 with \texttt{GephiForR}'s}
  \centering
  \begin{subfigure}{0.55\textwidth}
    \centering
    \includegraphics[width=\linewidth]{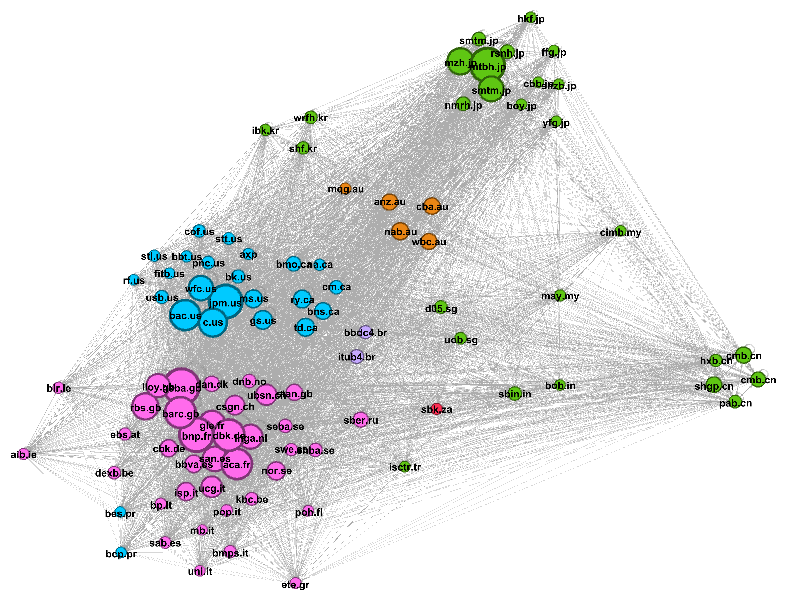}
    \subcaption{ForceAtlas2 in Gephi}
    \label{Gephi color}
  \end{subfigure}%
  \begin{subfigure}{0.44\textwidth}
    \centering
    \includegraphics[width=\linewidth]{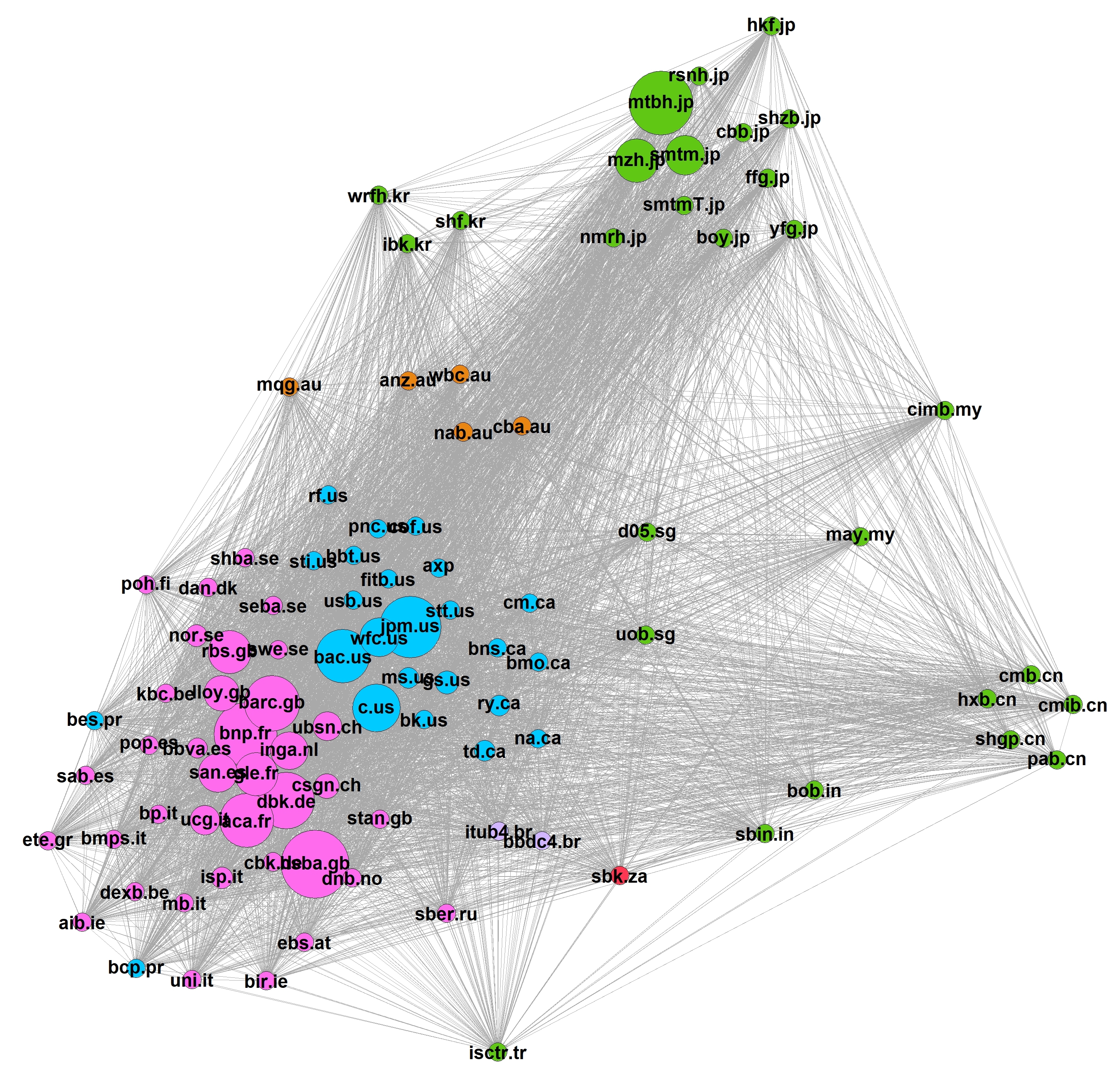}
    \subcaption{ForceAtlas2 in \texttt{GephiForR}}
    \label{GephiForR color}
  \end{subfigure}
  \label{Demirer bank plot}
    \caption*{\footnotesize Note: The left graph shows the ForceAtlas2 layout generated in Gephi while the right graph shows the layout resulting from \texttt{GephiForR}'s implementation of ForceAtlas2. The plots are largely similar while still having some differences due to initial random positions. Node colors are determined by continent, with blue being North America, purple being South America, pink being Europe, orange being Australia, green being Asia, and red being Africa.}
\end{figure}

I include a comparison of the layout generated with Gephi and that with ForceAtlas2. As described in Section \ref{FA2 main feature}, I follow the Demirer et al. example of estimating global banking connectedness via stock return volatility data. The node naming convention indicates the bank and country in the form ``bank.country." For instance, J.P. Morgan is ``jpm.us" in Figure \ref{Demirer bank plot}, and I color nodes by continent.

The plot below is produced via the series of functions above, in that order and exactly as specified. The only changes to note are that I do not scale the graph or pass in edge color: after creating the graph, I simply calculate the network layout with ForceAtlas2, rotate the layout to align with Gephi's output, assign node colors, and plot the diagram below via \texttt{easyplot}, as laid out in the above code examples.

While there are still some differences between the two plots, as is apparent, the plots are quite similar. Importantly, the clustering aligns between both plots: both ForceAtlas2 in Gephi and in \texttt{GephiForR} have clear clustering between pink and blue nodes,  distinct clusters of orange and purple nodes, and accurate clusters of green nodes as it registers differences between Japanese, Chinese, Korean, Indonesian, Singaporean, and Malaysian banks, all of which have unique clusters in both Figures \ref{Gephi color} and \ref{GephiForR color}. 

The variation in node position that does occur is largely due to the randomness of the initial positions. As described above, ForceAtlas2 in both Gephi and \texttt{GephiForR} assigns initial random positions, and nodes converge across several iterations. There is thus no one correct layout, and each recalculation of the layout will give different node positions. This means that the variation apparent in the plots above (such as the larger distance between the pink and blue nodes in Figure \ref{Gephi color} compared to Figure \ref{GephiForR color}) disappears while other variation is introduced, simply depending on initial node positions. In short, Figures \ref{Gephi color} and \ref{GephiForR color} should not perfectly align because the initial positions of the nodes are random: even with the same algorithm, the underlying randomness in assignment still persists, resulting in slightly different layouts. Yet, the key features\textemdash namely the distinct clustering\textemdash persist across random initial positions, providing a check of consistency. 

\begin{figure}[H]
    \centering
    \caption{An Alternate Implementation of ForceAtlas2 in R}
    \includegraphics[width=0.4\linewidth]{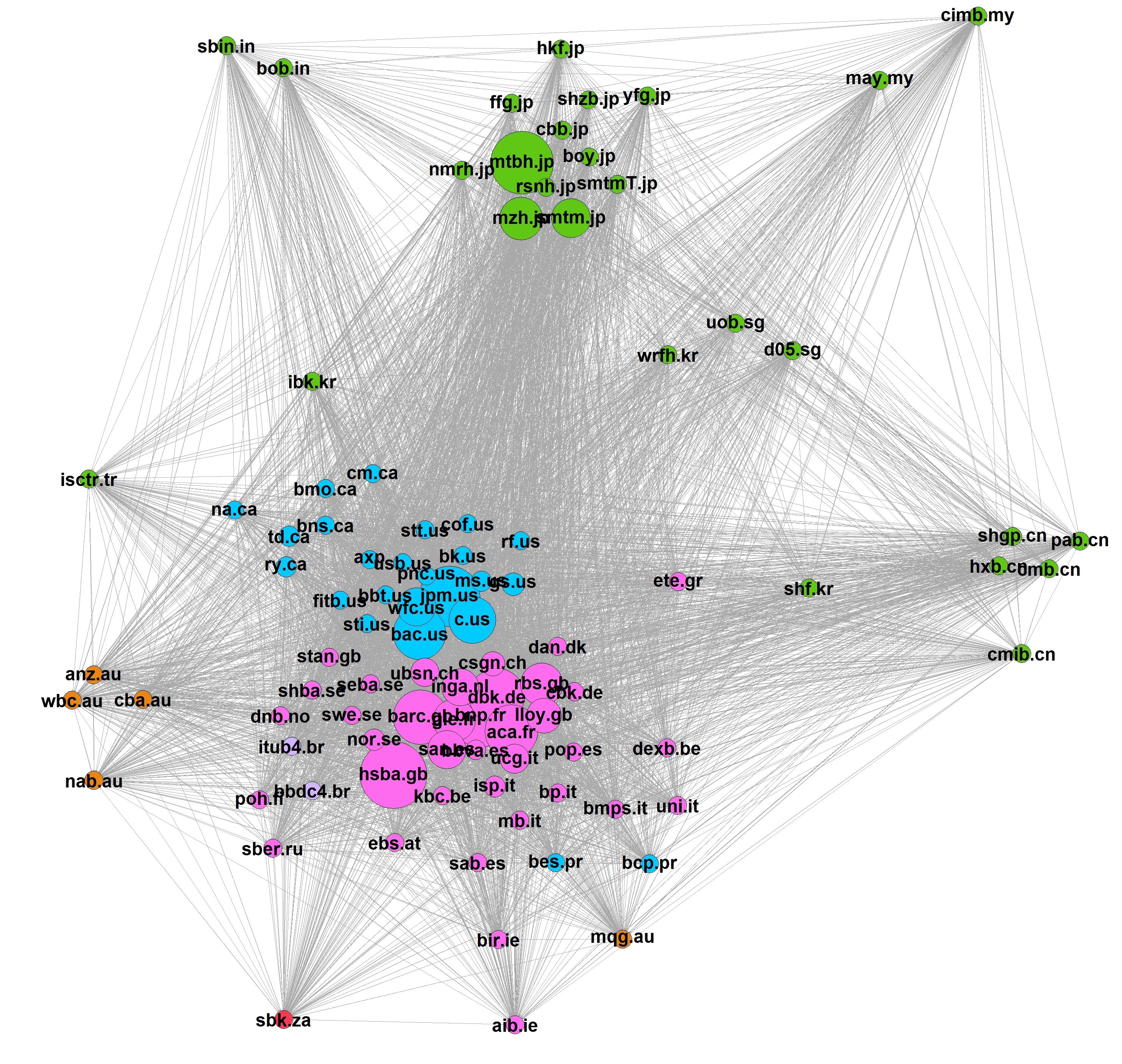}
    \subcaption{Klockiewicz and Álvarez's ForceAtlas2 in R}
    \label{fig:other alg}
    \caption*{\footnotesize Note: The above graph shows the ForceAtlas2 layout generated by Klockiewicz and Álvarez's ForceAtlas2 algorithm for R. As is apparent, the results are quite different from those in Figure \ref{Demirer bank plot}. Node colors are determined by continent, with blue being North America, purple being South America, pink being Europe, orange being Australia, green being Asia, and red being Africa.}
\end{figure}

These results are a marked improvement over other attempts to bring ForceAtlas2 to R; for instance, one implementation by Bazyli Klockiewicz and Adolfo Álvarez available on GitHub yields the result shown in Figure \ref{fig:other alg} (\cite{klockiewicz_forceatlas2_2016}). 

As is apparent, the Klockiewicz and Álvarez implementation still preserves clustering among pink and blue nodes and segments Japanese and Chinese firms (green), yet it particularly struggles with firms not in a large cluster. For instance, the small clusters of Korean firms (green) and Australian firms (orange) apparent in Figures \ref{Gephi color} and \ref{GephiForR color} are broken up in Figure \ref{fig:other alg}. Similarly, other firms without many in their clusters (such as Indonesian and Malaysian firms, both green) are pushed much further out on the periphery. This is because Klockiewicz and Álvarez's attractive force calculations are different than those in ForceAtlas2, and they also have a different speed calculation than Jacomy et al.'s Gephi. The result is that the average shape of the graph after several iterations is very different than that resulting from Gephi (and \texttt{GephiForR}). These differences in calculation method are also particularly limiting for time series visualization where the previous period's layout is passed in as the initial position for the current period: in this Klockiewicz and Álvarez implementation, the clusters disappear, and nodes tend to converge towards a Fruchterman-Reingold style layout.

\begin{figure}[H]
  \caption{Comparing Gephi's ForceAtlas2 with \texttt{GephiForR}'s, for Time Series Data}
  \begin{subfigure}{0.5\textwidth}
    \centering
    \includegraphics[width=\linewidth]{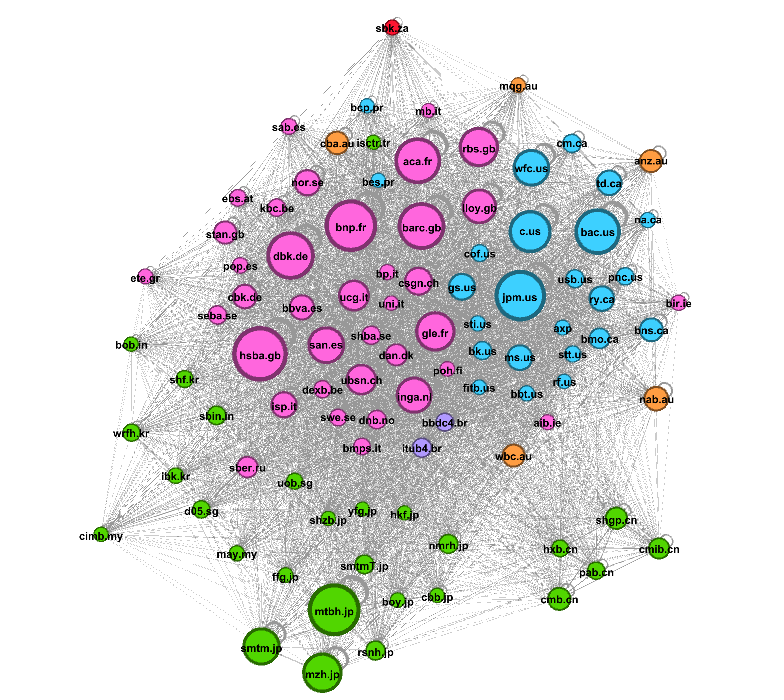}
    \subcaption{ForceAtlas2 in Gephi, Day 1}
    \label{Gephi color rw1}
  \end{subfigure}%
  \begin{subfigure}{0.52\textwidth}
    \centering
    \includegraphics[width=\linewidth]{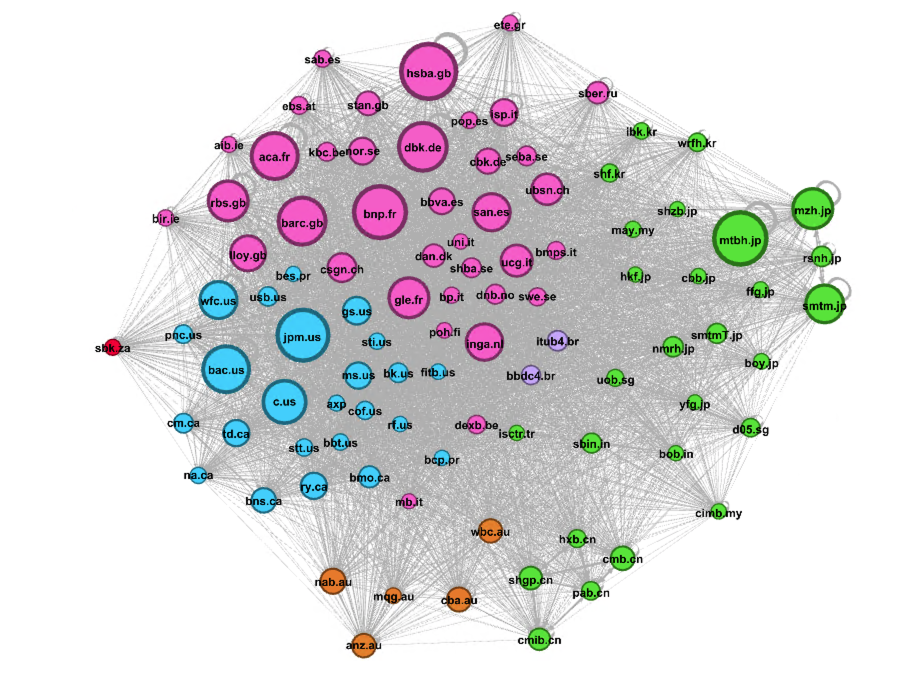}
    \subcaption{ForceAtlas2 in Gephi, Day 2}
    \label{Gephi color rw2}
  \end{subfigure}
    \centering
    
  \begin{subfigure}{0.49\textwidth}
    \centering
    \includegraphics[width=\linewidth]{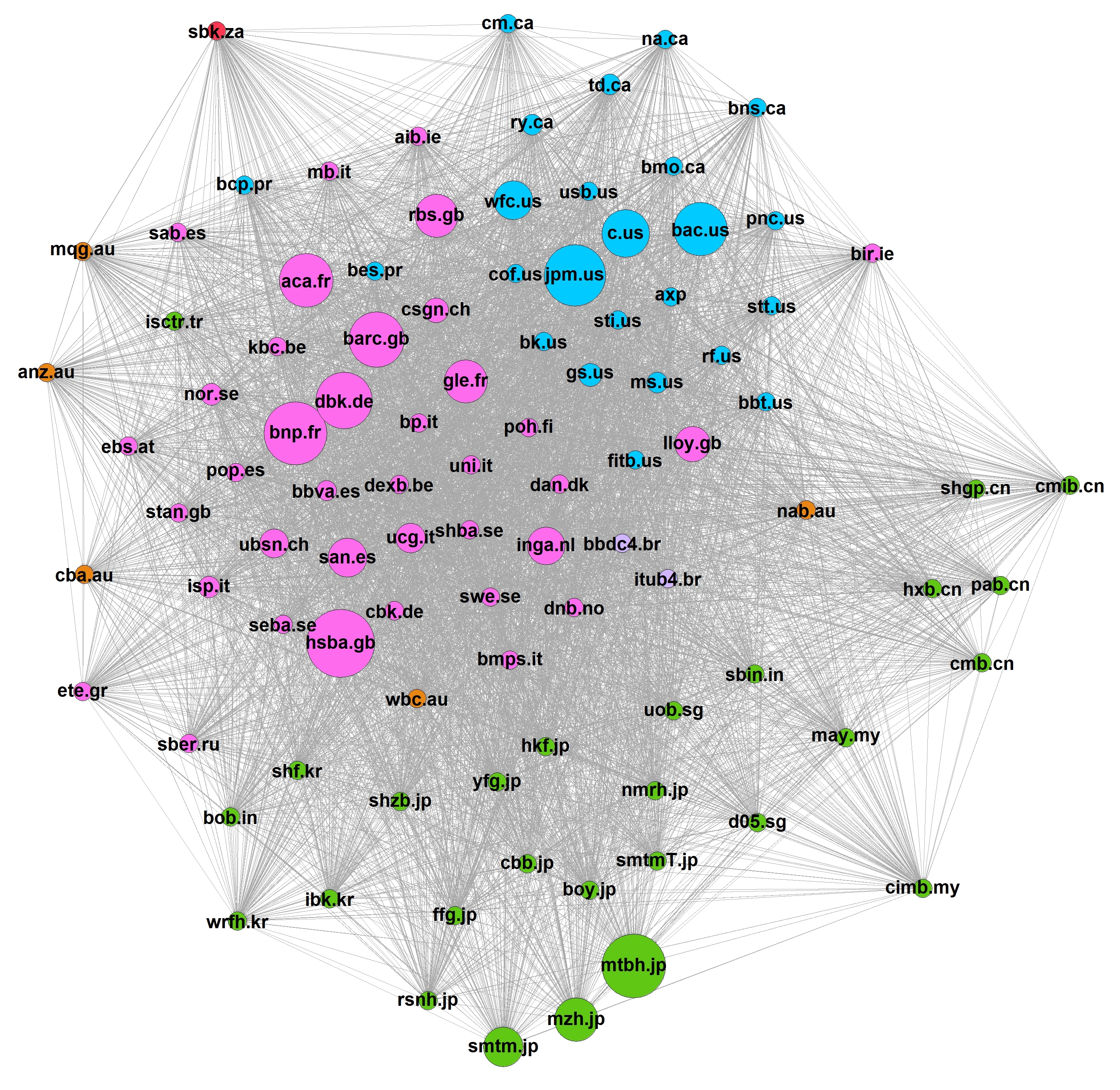}
    \subcaption{ForceAtlas2 in \texttt{GephiForR}, Day 1}
    \label{Gephiforr color rw1}
  \end{subfigure}%
  \begin{subfigure}{.51\textwidth}
    \centering
    \includegraphics[width=\linewidth]{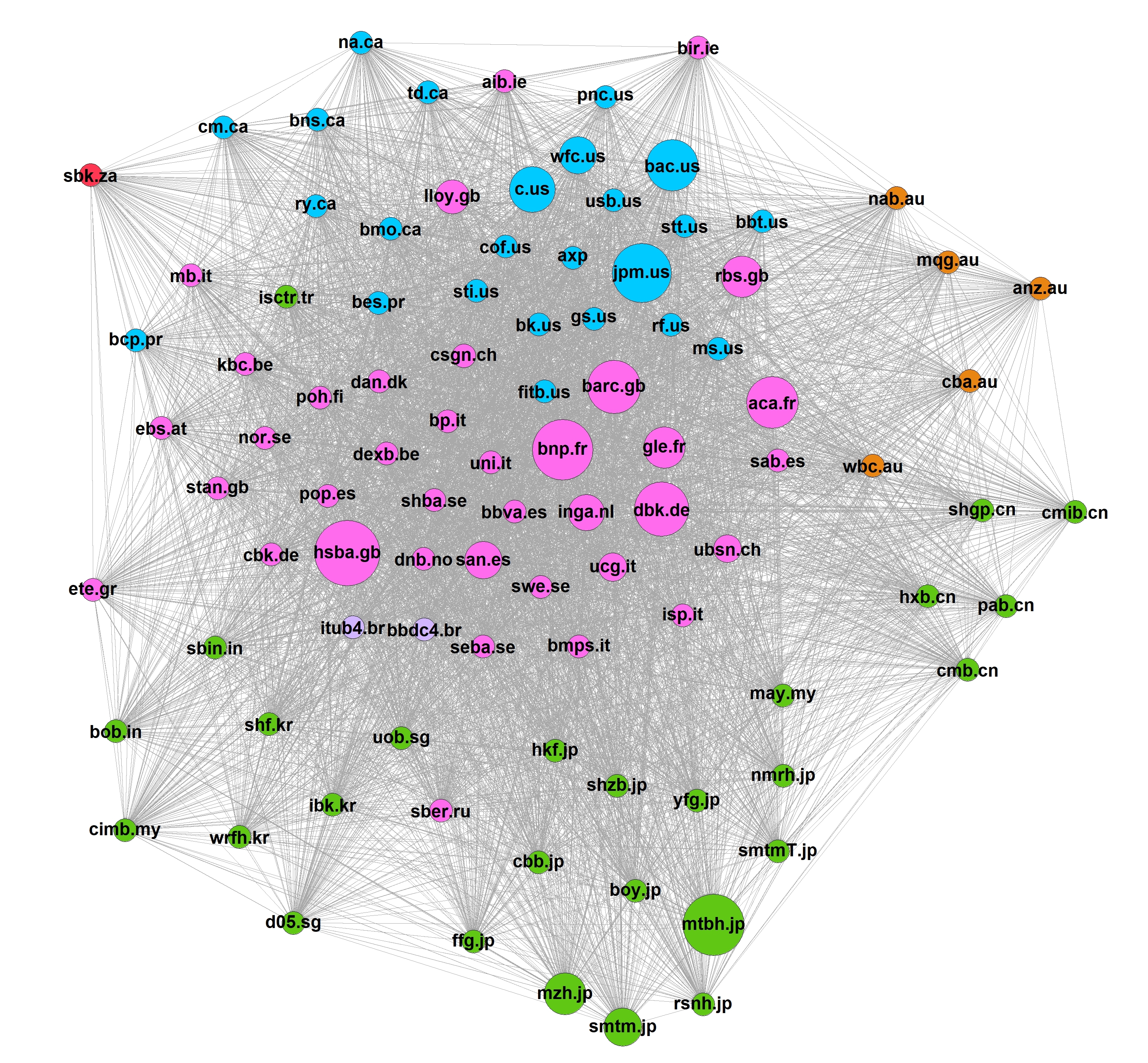}
    \subcaption{ForceAtlas2 in \texttt{GephiForR}, Day 2}
    \label{Gephiforr color rw2}
  \end{subfigure}
  \label{Demirer bank plot rw}
    \caption*{\footnotesize Note: The top row shows two networks generated with ForceAtlas2 in Gephi, with the upper left plot depicting the network for the 100-day period ending on August 22, 2008 while the upper right graph shows the 100-day window ending on the next business day, August 25, 2008. The bottom row illustrates plots generated by \texttt{GephiForR}'s ForceAtlas2 layout algorithm and plotting tools. The bottom left plot depicts the network for the 100-day window ending on August 22, 2008, and the bottom right plot illustrates the 100-day window ending on the next business day, August 25, 2008. As before, node colors are determined by continent, with blue being North America, purple being South America, pink being Europe, orange being Australia, green being Asia, and red being Africa.}
\end{figure}

I also provide an example of \texttt{GephiForR}'s expanded capabilities for plotting time series data. Again utilizing the Demirer et al. (2018) bank stock volatility data, I plot two sequential 100-day periods leading up to the financial crisis. These are ``rolling window" estimations, wherein 99 days of the sample overlap. As the window ``rolls" forward, the last day is added, and the first day is dropped. The networks below thus depict the 100 business days leading up to August 22, 2008 in Figures \ref{Gephi color rw1} and \ref{Gephiforr color rw1}, while Figures \ref{Gephi color rw2} and \ref{Gephiforr color rw2} illustrate the 100-day window ending in August 25, 2008. Thus, there should be a strong amount of stability between these plots from consecutive time periods, and differences should be easy to track in the network.

As is apparent, there is significantly more consistency across the plots generated with \texttt{GephiForR} than those resulting from Gephi itself. As previously described, this is because Gephi randomly assigns initial positions each time it reads a new network in. While it allows one to fix the position of a node or set of nodes, doing so for several periods worth of time series data would not only prove challenging and time consuming, but would effectively mean that the network rotates and cycles around the fixed node(s), which should have their own positions dynamically updated. Comparing the plots in \ref{Gephi color rw1} and \ref{Gephi color rw2}, changes in relative position are hard to detect simply because nearly every part of the network moves. 

In allowing for previous layouts to be passed in as initial positions, \texttt{GephiForR}, on the other hand, allows more stability between plots. Comparing the layouts on the different days (\ref{Gephiforr color rw1} and \ref{Gephiforr color rw2}) is now much easier because the nodes tend to remain closer to their initial positions, allowing differences in force to be detected much more easily.

Thus, following Gephi's algorithm, \texttt{GephiForR} provides convenient replication of ForceAtlas2 in R, limited only by the randomness of initial node position. Further, its built-in plotting functions allow one to replicate Gephi's features with ease, supporting and advancing beyond Gephi's capabilities.

\section{Efficiency and Scale} \label{efficiency and scale}

Importantly, one of the key limitations of \texttt{GephiForR} is R's own capabilities. Java excels in parallel computing, where a computation can be divided into subproblems, executed in separate threads, and then reaggregated. With this ``fork/join framework," Java can be quite efficient and quick, even on computers with limited memory and processor cores (\cite{noauthor_parallelism_2022}). Gephi is thus quite efficient at formatting and displaying large networks. On the other hand, R's baseline is single-thread processes. In R, multithreading is slow due to the overhead of data transfer to and from new threads, and it often depends on the operating system (for instance, Windows vs. Unix-based systems). Requiring longer times to copy information back and forth across threads, R's single-thread computations can be markedly faster than its parallelized ones depending on the speed of the computations conducted across the threads, relative to the amount of time spent copying the information across threads. 

Because of these features, my attempts to implement a parallelized version of ForceAtlas2 in \texttt{GephiForR} (\texttt{layout.forceatlas2}) only slowed down computation, increasing computation time by a factor of 1.5. The ForceAtlas2 algorithm included in \texttt{GephiForR} is thus a single-thread computation without parallelization. I display runtimes for the algorithm with computational time on the y-axis and the number of nodes on the x-axis in Figure \ref{fig:runtime} below, with each line representing different numbers of iterations. Times were calculated and averaged across 10 iterations on a Windows computer with a 4-core processor and 8 GB of RAM. Confidence intervals are shown via the vertical bars, though they are relatively narrow. 

\begin{figure}[H]
\caption{Computational Time by Number of Nodes and Iterations, with Confidence Intervals}
    \label{fig:runtime}
    \centering
    \includegraphics[width=.75\linewidth]{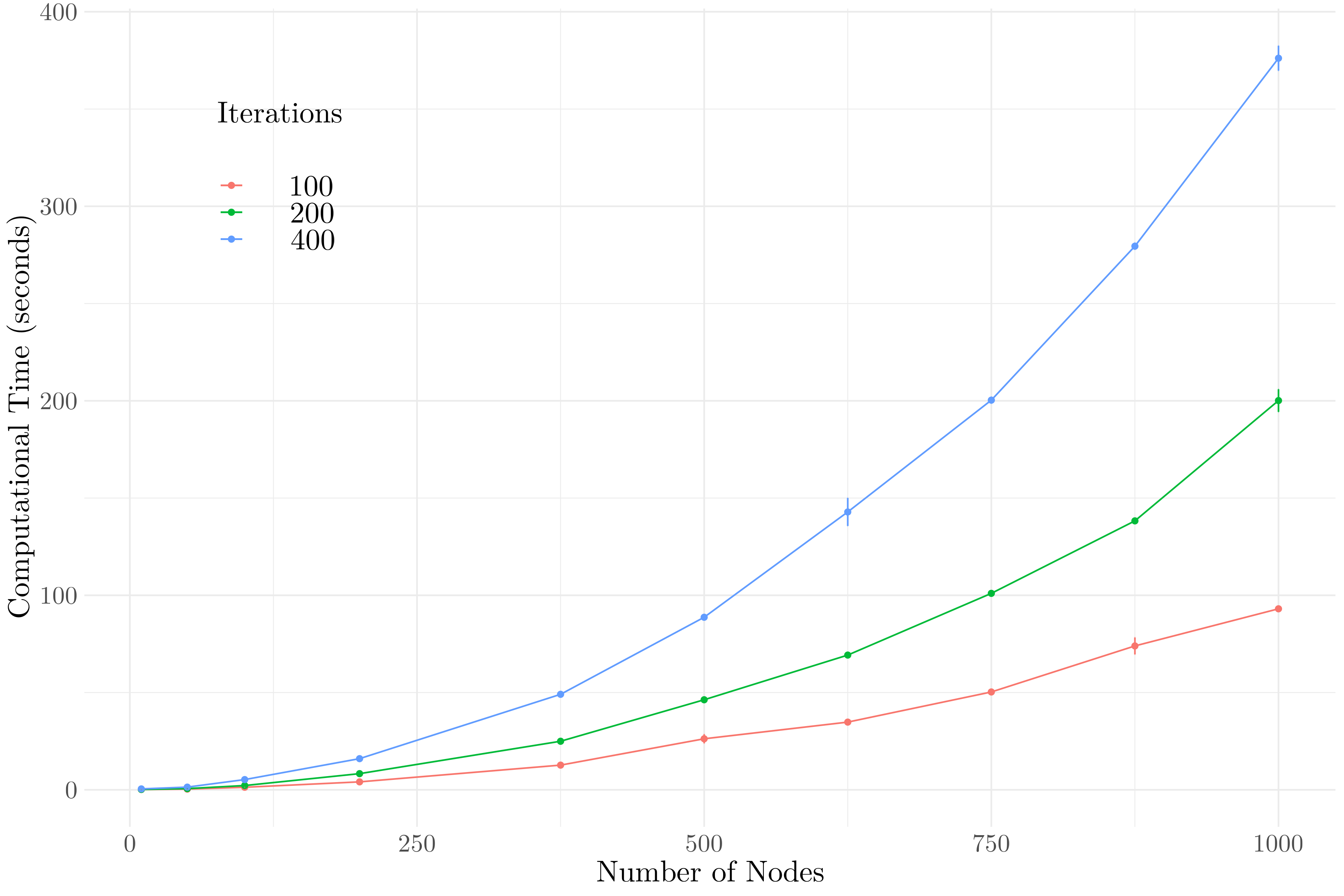}
        \subcaption{ForceAtlas2 in \texttt{GephiForR}}
        \caption*{\footnotesize Note: Computational time in seconds is on the y-axis, and the number of nodes is on the x-axis. The red line illustrates 100 iterations, the green reflects 200 iterations, and the blue captures 400 iterations. Confidence intervals are shown at the points calculated and are represented as vertical bars, but they are relatively narrow, as shown.}
\end{figure}

As is apparent above, computational time grows parabolically with the number of nodes and iterations. For smaller networks, \texttt{GephiForR}'s ForceAtlas2 is quite quick and efficient, consistently taking less than a minute; even for 1000 nodes, computation time with 100 iterations is only around 90 seconds. Yet, realistically, the larger the number of nodes in the network, the more iterations will likely be needed for convergence to occur. One would thus expect \texttt{iterations = 400} or more for networks of 1000 nodes, implying longer computation times. 

Thus, because of R's computational abilities, \texttt{GephiForR}'s ForceAtlas2 is best suited for smaller networks ($<1000$ nodes), but it can handle more if there is sufficient computer memory.\footnote{Networks of up to 5000 nodes were able to be processed on a Windows 4-core computer with 8 GB RAM, although computation time was correspondingly much lengthier.} Further, visualization of networks larger than 1000 nodes may also be difficult with R's static graphing structure; while one can click-and-drag the network to examine it in Gephi, it is not possible to do so in R. Gephi is thus the go-to tool for very large networks, simply because it has the computing power to quickly calculate ForceAtlas2 in a highly efficient, parallelized way.

\section{Discussion} \label{conclusion}
Bringing simple and efficient Gephi-style plotting to R, the \texttt{GephiForR} R package implements Gephi's key layout algorithm and its extensions, as well as other useful Gephi-style graph manipulation tools like rotation and plotting. With its node and edge color assignments, as well as its plotting function, \texttt{GephiForR} also provides convenient tools for graph visualization. Importantly, the package and its functions are designed to be accessible to those with little coding background, displaying informative error messages and providing instructional information. 

\texttt{GephiForR} advances past Gephi's original abilities to visualize time series data, allowing previously calculated layouts to be passed in as initial positions, and it therein helps anchor nodes across sequential periods. However, its chief limit is its ability to quickly and efficiently calculate ForceAtlas2 on large networks of 1000 nodes and over. 

Future work on \texttt{GephiForR} will focus on extending the features implemented, creating new features as Gephi continues to evolve, and other modifications depending on user demand. Importantly, computational efficiency is an ongoing focus, especially given \texttt{GephiForR}'s limited ability to handle large networks. Future versions of \texttt{GephiForR} will ideally be faster and more efficient in calculating ForceAtlas2.

\begin{CJK*}{UTF8}{bsmi} 
\printbibliography
\end{CJK*}

\end{document}